\documentclass[prd,twocolumn,superscriptaddress,floatfix,amsmath,amssymb,amsfonts]{revtex4-1}

\usepackage{float} 

\usepackage{comment}
\usepackage[normalem]{ulem}
\usepackage[english]{babel}
\usepackage{graphicx}
\usepackage{dcolumn}
\usepackage{bm}
\usepackage{blindtext}
\usepackage{verbatim}
\usepackage{mathrsfs}
\usepackage{amsmath}
\usepackage{cancel}
\usepackage{physics}
\usepackage{epstopdf}
\usepackage{mathtools}
\usepackage{color}

\newcommand{\ii}{\mathrm{i}}

\usepackage[usenames,dvipsnames]{pstricks}
\usepackage{epsfig}
\usepackage{pst-grad} 
\usepackage{pst-plot} 

\usepackage{hyperref}

\usepackage{verbatim}

\begin{document}

\title{General Relativistic Quantum Optics: \break Finite-size particle detector models in curved spacetimes}

\author{Eduardo Mart\'{i}n-Mart\'{i}nez}
\email{emartinmartinez@uwaterloo.ca}
\affiliation{Institute for Quantum Computing, University of Waterloo, Waterloo, Ontario, N2L 3G1, Canada}
\affiliation{Department of Applied Mathematics, University of Waterloo, Waterloo, Ontario, N2L 3G1, Canada}
\affiliation{Perimeter Institute for Theoretical Physics, Waterloo, Ontario, N2L 2Y5, Canada}

\author{T. Rick Perche}
\email{trickperche@perimeterinstitute.ca}
\affiliation{Perimeter Institute for Theoretical Physics, Waterloo, Ontario, N2L 2Y5, Canada}

\affiliation{Instituto de F\'{i}sica Te\'{o}rica, Universidade Estadual Paulista, S\~{a}o Paulo, S\~{a}o Paulo, 01140-070, Brazil}

\author{Bruno de S. L. Torres}
\email{bdesouzaleaotorres@perimeterinstitute.ca}
\affiliation{Perimeter Institute for Theoretical Physics, Waterloo, Ontario, N2L 2Y5, Canada}

\affiliation{Instituto de F\'{i}sica Te\'{o}rica, Universidade Estadual Paulista, S\~{a}o Paulo, S\~{a}o Paulo, 01140-070, Brazil}

\begin{abstract}

We propose a fully covariant model for smeared particle detectors in quantum field theory in curved spacetimes. We show how effects related to accelerated motion of the detector and the curvature of spacetime influence the way different observers assign an interaction Hamiltonian between the detector and the field. The fully covariant formulation explicitly leaves the physical predictions of the theory invariant under general coordinate transformations, hence providing a description of particle detector models (e.g., Unruh-DeWitt detectors, models for the light-matter interaction, etc) that is suitable for arbitrary trajectories in general spacetime backgrounds.

\end{abstract}

\maketitle

\section{Introduction}    
    The notion of measurement in quantum field theory (QFT) is much more delicate than its (already full of subtleties) counterpart in non-relativistic quantum mechanics. Projective measurements on localized field observables, for example, raise issues regarding locality and causality, thus being difficult to reconcile with the very pillars of quantum field theory \cite{Sorkin, dowker2011useless, Benincasa_2014}. Circumventing these problems while still being able to perform measurements that extract information from a quantum field is achieved by introducing the idea of particle-detector models \cite{DeWitt,Unruh-Wald}, where a particle detector is conceived as a localized, first-quantized system with internal degrees of freedom that couple to a local field observable. 
    
    Particle-detector models provide a very appealing operational  framework  from the point of view of how one intuitively understands the notion of measurement of a local observable in field theory. From the retinas of our eyes to solid state sensors at the LHC, we never measure a quantum field observable other than by coupling something to it. 
    
    Furthermore, detectors are very appealing as well from a fundamental perspective, since they allow us to  extract information locally from a quantum field. In this context, particle-detector models have proven to be a powerful tool to understand well-known effects in QFT such as Unruh and Hawking effects \cite{Unruh1976, Sciama1977}. What is more, particle detector models have been shown to capture the fundamental features of the interaction of atoms with the quantum electromagnetic field, hence also providing a very natural way to bridge fundamental theories with experimental setups relevant to fields such as quantum optics and relativistic quantum information.
    
    Although most of the initial applications of detector models in the previously discussed examples dealt with pointlike detectors, there has been an increasing interest in issues related to smeared detectors. One of the original motivations to smear particle detectors was to regularize UV divergences associated to purely pointlike interactions, but it is also important to note that the notion of smeared detectors  naturally arises in the context of the light-matter interaction, since atoms themselves are not pointlike objects. Smeared detectors are also ideal from the point of view of algebraic QFT, where the local algebras of observables (to which smeared particle detectors happen to couple) are considered to be the fundamental object of the theory \cite{Chris}.
    
    In the common approach to smeared detectors in the literature, the Hamiltonian of smeared particle detectors is prescribed in the detector's `centre of mass' frame. The usual prescription is such that the local profile of the interaction happens in the simultaneity surfaces to the centre of mass' world line. However, there is a problem with this approach, which is that the time evolution operator obtained through this method cannot consistently be made independent of the coordinates chosen to describe the interaction, a requirement of covariance we would like to have in any physical description of detector-field interactions.
    
     One way of thinking about this problem while keeping an explicitly covariant approach is to write the time evolution operator in terms of a spacetime integral of a Hamiltonian density, much in the spirit of what is done when assigning interactions in field theory. 
    
   The covariance of physical predictions of particle-detector models in arbitrary trajectories (in flat spacetimes) was addressed in \cite{eduardo}. In the present paper, we aim to generalize this formalism to scenarios where strong accelerations and gravitational effects may play an important role (i.e., in curved spacetimes and very non-inertial trajectories). We outline a systematic procedure by which one could identify time evolution with respect to different notions of time translation in curved spacetimes, and use that to formulate an explicitly covariant coupling between detectors and quantum fields. In this formulation, the predictions of the theory are automatically invariant under general changes of reference frame. 
   
   We will also explicitly evaluate the error that one incurs when using the non-covariant approach as a function of the relevant lengthscales of the detector smearing, showing in what regimes the non-covariant model (used in the detector's centre of mass proper frame) is a good approximation for the fully covariant predictions.  In particular, we discuss how, for a single detector, the non-covariant approach is a good approximation in most scenarios. However, this is not as simple when considering systems of several detectors, where the covariant formulation becomes especially crucial. This is the case in various problems of interest to the field of relativistic information such as e.g., entanglement harvesting \cite{Valentini1991,Reznik1,reznik2,Pozas-Kerstjens:2015,Nick,Olson2011,Olson2012,Salton:2014jaa,Pozas2016,Nambu2013,Kukita2017,Kukita20172,Ng2,Henderson2018,Henderson2019,Petar,Cong2019,Brown1,Brown2,Retzker2005,Sabinprl,Farming} or classical and quantum communication through quantum fields \cite{ClicheKempfD,Jonsson1,Jonsson2,Jonsson3,Jonsson4,eduardo,Katja,Landulfo,Funai,Shockwaves,simidzija2019transmission}.
    

\section{Time Translations in General Spacetimes}
    
    The canonical interpretation of the Hamiltonian assigns to it the status of generator of time translations. In a relativistic scenario, however, one has to be aware of the different notions of time translations for observers in different states of motion -- a problem which, for systems of particle detectors, has already been addressed in e.g., \cite{Brown:2014en, eduardo}. The guiding line in such situations, where it may not be clear with respect to which time coordinate you should be evolving the system, is to interpret the time evolution operator as an operator that takes the state of a system from a spacelike surface $\mathcal{E}_1$ and evolves it to another spacelike surface $\mathcal{E}_2$, and then the time evolution operator is obtained from integrating a Hamiltonian density in the spacetime region between $\mathcal{E}_1$ and $\mathcal{E}_2$ \cite{eduardo}.

    Often, in many physical scenarios there is some underlying criteria to choose the spacelike surfaces between which the system is evolving in a clear and unambiguous way, directly associating them to to the surfaces of constant $t$ (where $t$ is some relevant time coordinate to the physical setup  employed). However, when one has more than one relevant notion of time translation in a given problem---and perhaps even different Hamiltonians generating translations with respect to different timelike directions--- the situation can become more complicated.
    
    Take for example a particle detector that is not comoving with a the quantization frame of a quantum field theory (QFT). In that case it will be necessary to relate time translations with respect to the detector's proper time and the time translations with respect to the coordinate time in the frame where the field quantization is performed. Things are even more complicated when the detector is spatially smeared (something necessary to e.g., model the light-matter interaction or avoid pervasive divergences associated with pointlike interactions) in which case it may not even make sense to talk about the `detector's proper time'.

    In quantum field theory in flat spacetimes, canonical quantization of a free field relies heavily on one being able to describe the classical dynamics of the field in terms of a complete set of solutions to the equations of motion (Klein-Gordon, Maxwell, etc.). An analogous approach to the quantization of a free field in a curved background $\mathcal{M}$ is possible, provided that the background spacetime allows for a well-defined initial-value formulation to the classical equations of motion for the field. Whether this is possible is determined by the causal structure the particular spacetime. 
    
    To formulate this problem, let us recall some relevant definitions regarding causal structure. A \textit{causal curve} is a curve that is either timelike or null; an \textit{achronal surface} is a closed surface $\Sigma \in \mathcal{M}$ such that no two points in $\Sigma$ can be connected by a timelike curve; the \textit{domain of dependence} of a surface $\Sigma$ (denoted by $D(\Sigma)$) is the set of all points  $p \in \mathcal{M}$ such that every inextendible causal curve passing through $p$ intersects $\Sigma$; an achronal surface $\mathcal{E} \subset \mathcal{M}$ is said to be a \textit{Cauchy surface} if $D(\mathcal{E}) = \mathcal{M}$; and finally, a spacetime is said to be \textit{globally hyperbolic} if it possesses a Cauchy surface. Global hyperbolicity is a sufficient condition for the well-posedness of an initial value formulation for the equations of motion, and therefore throughout the present article we will always be assuming this condition on the background spacetime.
    
    In this situation, one can conceive a Hamiltonian that is to be interpreted as the generator of time translations from one Cauchy surface to another. (For more details on relevant concepts for discussing causal structures of spacetimes, see  \cite{Wald1}; for further discussion on a detailed procedure to quantize a free scalar field in a globally hyperbolic spacetime, see \cite{Wald2}).

    On the other hand, assume that there is an observer following a time-like trajectory $\mathsf{x}(t)$ where $t$ is a time-like coordinate. It can then be reparametrized by its proper time $\tau$, yielding a parametrization $\mathsf z(\tau)=\mathsf x(t(\tau))$. The constant time surfaces associated to such an observer at proper time $\tau_0$ are defined by the exponential of all vectors in $T_{\mathsf z(\tau_0)}\mathcal{M}$ orthogonal to its four velocity $u^\mu(\tau_0)$. Let us call those surfaces $\Sigma_\tau$. There is a natural coordinate system to describe these surfaces that is locally defined around such trajectory: the Fermi normal coordinates $\bar{\mathsf x} = (\tau,\bm{ \bar{x}})$. These coordinates are such that for each fixed $\tau_0$, $(\tau_0,\bm{ \bar{x}})$ defines local coordinates for $\Sigma_{\tau_0}$ around the observer's position $\mathsf z(\tau)$. Note that, by construction, we have $\mathsf u(\tau) = \partial_\tau$ along the curve.
    
    According to the discussion above, the Hamiltonian that generates time translations with respect to the observer's proper time will induce time evolution between the spacelike surfaces $\Sigma_{\tau_1}$ and $\Sigma_{\tau_2}$. If we want to compare the time evolution provided by the Hamiltonian associated to this observer with the one provided by the quantization coordinates, we must assume that for some value of $\tau_1$ and $t_1$ we have $\mathcal{E}_{t_1} = \Sigma_{\tau_1}$, so that we evolve the system from the same spacelike surface. We also assume that $\mathcal{E}_{t_2} = \Sigma_{\tau_2}$ for final times $t_2$ and $\tau_2$, in such a way that the final state of the field is defined in the same region of spacetime after such an evolution. In the present paper we shall assume that these conditions hold true asymptotically.
    
    The simplest possible scenario where these subtleties are important is when one talks about time reparametrization invariance. To illustrate this point, let us consider time evolution in quantum mechanics, where we know that time evolution with respect to a given coordinate time $t$ can be related to the Hamiltonian according to the Schr\"odinger equation:
    \begin{equation}
        \hat H^t(t)\ket{\psi} = \ii\frac{\dd}{\dd t}\ket{\psi}.
    \end{equation}
    This means that the Hamiltonian with respect to another time parameter $\tau$ (say, for example, the proper time of an observer along a curve $z(\tau)$) is expected to satisfy
    \begin{equation}
        \hat H^\tau(\tau)\ket{\psi} = \ii \frac{\dd}{\dd \tau} \ket{\psi},
    \end{equation}
    which gives explicitly the relationship between $H^\tau$ and $H^t$ derived from time reparametrization:
    \begin{equation}\label{hTauHT}
        \hat H^\tau = \frac{\dd t}{\dd \tau} \hat H^t = u^t\hat H^t.
    \end{equation}
    The derivative of $t$ with respect to $\tau$ was written in terms of the $t$ component of the four-velocity to emphasize that having the curve is enough to relate the Hamiltonians in this case. In more complicated cases, in which different notions of time translations are not exclusively due to time reparametrization, but to general spacetime coordinate transformations, and when we consider smeared particle detectors coupled to quantum fields in curved spacetimes, the calculation is more involved; we will explicitly develop the procedure below.
    
\section{The Detector-Field System Hamiltonian}

    Let us then formulate our model for describing the coupling between a free scalar field $\phi$ and an Unruh-DeWitt detector in a spacetime $\mathcal{M}$ equipped with a metric $g$.  The assumption of globally hyperbolic spacetime allows us to foliate it as ${\mathcal{M} = \bigcup_{t\in \mathbb{R}} \mathcal{E}_t}$, where each $\mathcal{E}_t$ is a Cauchy surface. This selects a timelike direction $\partial_t$, and therefore the surfaces $\mathcal{E}_t$ generalize the idea of ``constant time surfaces''. This allows us to introduce ``equal time'' commutation relations to canonically quantize a classical field.

For concreteness, we are going to focus on free scalar field quantization and on particle detectors that couple linearly to the field. This scenario is paramount in many studies in quantum field theory in curved spacetimes, relativistic quantum information, and in quantum optics, since scalar particle detector models such as the Unruh-DeWitt (UDW)~\cite{DeWitt,Unruh-Wald} detector have been proven to be good models for the light-matter interaction in most regimes~\cite{Pozas2016,eduardo}. The theory of a free scalar field in an arbitrary curved spacetime in $D=n+1$ dimensions is described by an action given by
\begin{equation}
    S[\phi] = \int \!\dd^{D}\mathsf x\: \sqrt{-g}\left(-\dfrac{1}{2}\nabla_{\mu}\phi\nabla^{\mu}\phi  - \dfrac{1}{2}m^2 \phi ^2\right),
\end{equation}
whose Euler-Lagrange equations give us the Klein-Gordon equation: $(\nabla_{\mu}\nabla^{\mu} - m^2)\phi = 0$.

Under the assumptions described above the free evolution of the field $\phi$ is given by a general solution that can be written as
\begin{equation}
     \phi(\mathsf x) = \int \dd^{n} \!\bm{k}\left(a^*_{\bm{k}} u_{\bm{k}}(\mathsf x)+a_{\bm{k}} u^*_{\bm{k}}( \mathsf x) \right),
\end{equation}
    where the functions $u_{\bm{k}}(x)$ form a complete set of solutions to the Klein-Gordon equation. Upon quantization, the amplitudes are promoted to creation and annihilation operators $\hat a^{\dagger}_{\bm{k}}$ and $\hat a_{\bm{k}}$ which will be then imposed to satisfying canonical commutation relations. Namely, we impose $
        [\hat a_{\bm{k}},  \hat a^{\dagger}_{\bm{k}'}] =  \delta^{(n)}(\bm{k} - \bm{k}')\openone $,
    with all other commutators vanishing.
    
    The UDW detector will be described as a two-level system that interacts with the field along a time-like trajectory $\mathsf z(\tau) = (t(\tau),\bm{z}(\tau))$ parametrized by proper time $\tau$. This means that the free Hamiltonian that generates translation with respect to the proper time of the centre of mass of the detector can be written as
    \begin{equation}\label{hD}
        \hat H_{d}^\tau = \Omega \hat \sigma^{+}\hat \sigma^{-},
    \end{equation}
    where $\Omega$ is the energy gap between ground and excited states of the detector as measured in the detector's proper reference frame \cite{eduardo}, and $\hat \sigma^{+}$ and $\hat \sigma^{-}$ are the usual SU(2) ladder operators. It is important to stress once again that such a Hamiltonian generates time translations with respect to the proper time of the detector, which can be related to the quantization coordinate time $t$ by the expression \eqref{hTauHT}.
    
    Now, the last ingredient to be included is the coupling between detector and field.
    One can define a monopole operator for the detector which in the interaction picture takes the form~\cite{eduardo}
    \begin{equation}
             \hat \mu(\tau) = e^{i\Omega \tau}\hat \sigma^+ +e^{-i\Omega\tau}\hat \sigma^-.
    \end{equation}
The interaction Hamiltonian between detector and field is, as usual, prescribed in the proper frame of the detector, which is well-defined for pointlike detectors~\cite{eduardo}:
    \begin{equation}
        \hat H_{I}^{\tau} = \lambda \chi(\tau)\hat \mu\,\hat \phi(\mathsf z(\tau)),
    \end{equation}
     where $\chi(\tau)$ is the \textit{switching function}, responsible for controlling the time duration and intensity of the interaction.

    \section{Smeared Detectors in curved spacetimes}
    
    The interaction Hamiltonian defined in the previous section models the detector as a point-like object that interacts with the field only along its trajectory, that is, only at a single point in space for every value of its proper time.  Pointlike interactions, however, are linked to unphysical divergences~\cite{Jorma} and, although sometimes acceptable approximations, pointlike systems are arguably  controversial choices to represent physically meaningful detectors coupling to quantum fields. Take for instance the light-matter interaction in quantum optics: atoms are reasonably localized objects that couple to the electromagnetic field in a finite region of space. Since  such interactions happen locally but  not only in a pointlike manner,  it is not enough to restrict ourselves to pointlike detectors. Also, from a fundamental point of view, a quantum field theory is described by its local algebras of observables, which are seen by many as the fundamental objects of the theory~\cite{Chris,Rechris,Halvorson}. These are all important motivations to look for a model that allows the detector to interact with elements of the local algebras of observables of a quantum field. In summary, we need to introduce \textit{smeared detectors}.
    
    If one were asked to write down a Hamiltonian that exhibits a coupling between detector and field with some spatial extension, perhaps the first straightforward way one could think of would be
    \begin{equation}
        \int_{\mathcal{E}'_{t'}} \!\!\! \dd^n \bm x' \sqrt{g'_{\mathcal{E}'}}\: \hat h_I(\mathsf{x}').
    \end{equation}
    where $\hat{h}_I(\mathsf{x}')$ is to be interpreted as a Hamiltonian weight  that describes the spatial profile of the interaction in the primed reference frame, and $\mathcal{E}'_{t'}$ is a spacelike hypersurface naturally adapted to the given frame. $g'_\mathcal{E'}$ is the determinant of the induced metric in these sheaves. One could naively expect the Hamiltonian to change according to expression \eqref{hTauHT} considering the fact that the space integral must now be performed in the spacelike surfaces $\mathcal{E}_t$, orthogonal to the new coordinate time. Therefore, it is only natural to assume that the Hamiltonian in the unprimed reference frame should be
    
    \begin{equation}
        \dv{t'}{t}\int_{\mathcal{E}_{t}} \!\!\!\dd^n \bm x\sqrt{g_{\mathcal{E}}}\: \hat h_I(\bm x).
        \label{noncov}
    \end{equation}    
    However, for reasons that will become clear in the following sections, this prescription can only be appropriate in the case of inertial detectors in flat spacetimes, not being a satisfactory one for smeared detectors in arbitrary trajectories in globally hyperbolic spacetimes since, as discussed below, it would yield incompatible predictions for different reference frames.

     In constructing a  fully covariant model for the interaction of field and detector, we will have to prescribe a Hamiltonian density $\hat{ \mathfrak{h}}_I(\bar{\mathsf{x}})$ that couples the field and the detector. Although there is no natural way of assigning such Hamiltonian density using non-relativistic quantum mechanics, first principle arguments will prescribe the form of this Hamiltonian density in the detector's centre of mass frame, while the  explicit covariance demands that we should unambiguously define how each observer extracts measurable quantities for the system.
    
 In what follows we will present the non-covariant model often used in the literature, analyze its drawbacks, then propose the fully covariant detector model described above, and finally explicitly compare the predictions of both models, discussing in what regimes keeping explicit covariance is fundamental and in what regimes the non-covariant model can be a good approximation.
    
       \subsection{Non-covariant prescription in the reference frame of the detector}\label{noncosec}
    
    Let us assume we have a detector which is smeared and whose centre of mass describes a trajectory $\mathsf z(\tau)$, where $\tau$ is the proper time of such curve. It is then useful to use Fermi normal coordinates around $\mathsf z(\tau)$. This coordinate system is defined as follows: for a fixed value $\tau_0$, we have the four-velocity of the curve given by \mbox{$\mathsf u(\tau_0) = \dot{\mathsf z}(\tau_0)$}, where the dot denotes derivatives with respect to $\tau$. The four-velocity is of course a normalized time-like vector. We can choose vector fields $\mathsf{e}_i$ in $\mathsf z(\tau_0)$ such that $\{\mathsf u(\tau_0),\mathsf{e}_i\}$ define an orthonormal basis of $T_{\mathsf z(\tau_0)}\mathcal{M}$ (the tangent vector space to spacetime at $\mathsf{z}(\tau_0)$. We then extend the basis $\mathsf{e}_i$ to the whole curve imposing that they are Fermi-Walker transported along it. We say a given vector field $v^\mu$ is Fermi-Walker (FW) transported if it satisfies the following differential equation:
    \begin{equation}
        u^\nu\nabla_\nu v^{\mu}=\left(v_{\nu} a^{\nu}\right) u^{\mu}-\left(v_{\nu} u^{\nu}\right) a^{\mu},
    \end{equation}
    where $a^\mu$ is the proper acceleration of the curve.
    
    It is worth mentioning that $u^\mu(\tau)$ is automatically FW transported and such transport allows us to extend the basis $\mathsf{e}_i$ along the curve $\mathsf{z}(\tau)$ ensuring that the extensions $\mathsf{e}_i(\tau)$ together with $\mathsf{u}(\tau)$ are still an  orthonormal basis for every $\tau$. We can thus write every vector orthogonal to $\mathsf{u}(\tau)$ in each of the tangent spaces $T_{\mathsf{z}(\tau)}\mathcal{M}$ as $\bar{x}^i \mathsf{e}_i(\tau)$, where $\bar{x}^i$ are real parameters. One can then define coordinates in the spacelike surfaces orthogonal to $u^\mu$ by exponentiating the vectors $\bar{x}^i\mathsf{e}_i(\tau)$. It is possible to show that such operation is well defined locally around the curve, and that this actually defines a coordinate system \cite{poisson}. Let $\Sigma_\tau$ stand for the spacelike hyper-surface generated by the $\mathsf{e}_i(\tau)$'s, that is, the rest space (hyperplane of simultaneity) with respect to the instantaneous four-position $\mathsf{z}(\tau)$. If the moving frame is associated with the trajectory of a particle detector, we could think of $\mathsf{z}(\tau)=(\tau,\bm 0)$ as the trajectory of its  centre of mass parametrized by proper time $\tau$.
    
    From this construction it follows that $\bm{\bar{x}} = \bar{x}^i$ define coordinates of the rest space relative to the trajectory $\mathsf{z}(\tau)$. We then define the Fermi normal coordinates around the curve by $(\tau,\bm{\bar{x}})$, which parametrize spacetime in an open set around the $\mathsf{z}(\tau)$. Furthermore, the properties of the exponential ensure that the proper distance between $\mathsf{z}(\tau)$ and a point with coordinates $\bar{x}^i$ in the sheave $\Sigma_\tau$ is given by $r = \sqrt{\bar{x}^i\bar{x}_i}$.
    
    We denote the components of the metric in Fermi normal coordinates by $\bar{g}_{\mu\nu}$, so that in each one of the sheaves $\Sigma_\tau$ we get an induced metric $\bar{g}_{ij}$. It is also important to remark that the Fermi normal coordinates are not orthogonal at every point. In fact, it is not even true that the vector $\partial_\tau$ is orthogonal to the surfaces $\Sigma_\tau$ at every point due to the presence of curvature. In Appendix \ref{expansion} we include the expansion of the metric up to second order on the distance from the points to the curve, where we see explicitly that curvature breaks such orthogonality condition. 
    
    Having such coordinate system naturally adapted to the rest space around the world line of the centre of mass of a particle detector, we can define an interaction Hamiltonian that takes into account a local interaction between the field and the detector as follows:
    \begin{equation}
        \hat H_I^\tau(\tau) = \lambda \chi(\tau)\hat{\mu}(\tau)\int_{\Sigma_\tau}\!\!\!\dd^n\bar{\bm x}\sqrt{\bar{g}_\Sigma}\: f(\bm{\bar{x}})  \hat \phi(\bar{x}),
        \label{Prescription1}
    \end{equation}
    where $f(\bm{\bar{x}})$ is called the \textit{smearing} function, and is responsible for determining the spatial profile of the interaction. The term $\dd^n\bar{x}\sqrt{\bar{g}_\Sigma}$ is  the induced volume element in the surface $\Sigma_\tau$, with $\bar{g}_\Sigma$ being the determinant of the spatial metric coefficients $\bar{g}_{ij}$. Just as in the point-like case, we could have made the physically reasonable assumption that the interaction Hamiltonian is prescribed in the reference frame of the detector, so that it generates time evolution with respect to the parameter $\tau$. This Hamiltonian is the one that has been extensively used in the literature of particle detectors (see, e.g. among many others, \cite{Schlicht, Jorma, eduardo}).
    
    It is then possible to calculate the time evolution operator, given by the time ordered exponential of the Hamiltonian
    \begin{equation}\label{uNonCov}
        \hat U = \mathcal{T}\exp\left(-\ii\lambda \int_{\mathbb{R}}\! \dd \tau\!\int_{\Sigma_\tau}\!\!\!\dd^n\bar{\bm x}\sqrt{\bar{g}_\Sigma}\:\chi(\tau)\hat\mu(\tau) f(\bm{\bar{x}})  \hat \phi(\bar{\mathsf x})\right),
    \end{equation}
    where the support of the switching function $\chi(\tau)$ encodes the possibly finite nature of the interaction.
    
   However, we see that this prescription explicitly breaks covariance. This can be seen by noting that the term in the exponential of the unitary evolution operator is not a covariant integral, in the sense that it cannot be written as an integral over an $D = n+1$ dimensional volume. This is because Fermi normal coordinates are orthogonal strictly on the trajectory of the centre of mass, but not around it. We see therefore that when there is strong curvature or acceleration, the prescription \eqref{Prescription1} may not be adequate. Discussing this in greater detail is the topic of the following section.

    \subsection{Covariant prescription of particle detector physics}
    
    We now present a fully covariant description of a smeared particle detector. As was seen in the previous section, what broke covariance was the fact that the unitary time evolution operator was defined as an integral in the detector's proper time in terms of the Hamiltonian seen in its reference frame. To solve this problem, we instead define a Hamiltonian density in a way that $U$ depends on an integral over a spacetime volume. This will yield a covariant formulation of this interaction. We then assume there is a \textit{spacetime smearing} compactly supported function $\Lambda(\mathsf x')$, which regulates both the spatial and temporal profile of the interaction. The interaction Hamiltonian weight can then be written as
    \begin{equation}
        \hat h_I(\mathsf x') = \lambda \: \Lambda(\mathsf{ x}') \hat\mu(\tau(\mathsf{x}')) \hat \phi(\bar{\mathsf x}'),
    \end{equation}
    so that the interaction Hamiltonian density can be written as \mbox{$\hat{\mathfrak{h}}_I = \hat h_I(\bar{\mathsf{x}})\sqrt{-\bar{g}}$.}
    
    The time evolution operator can thus be written as the exponential of the integral over the whole spacetime of $\hat{h}_I(\mathsf x)$:
    \begin{equation}\label{uFullCov}
        \hat{\mathcal{U}} = \mathcal{T}\exp\left(-\ii\int_\mathcal{M}\!\!\! \dd^D\mathsf x'\sqrt{-g'}\: \hat h_I(\mathsf x') \right),
    \end{equation}
    where $\bar{g}$ represents the determinant of the full metric in Fermi normal coordinates.
    It important to remark that the fact that $\Lambda(\mathsf x')$ is compactly supported ensures that the above integral is always taken in a compact region in spacetime, associated to the total time in which the interaction takes place and the spatial profile of the detector. If one wants to then write the time evolution operator as the time evolution generated by a Hamiltonian with respect to an arbitrary time coordinate $t'$, the natural prescription for such a Hamiltonian would be
    \begin{equation}\label{hCovGeneral}
        \hat{\mathcal{H}}^{t'}_I(t') = \int_{\mathcal{E}'_{t'}} \!\! \dd^n \bm{x}' \sqrt{-g'}\:\hat{h}_I(\mathsf{x}'),
    \end{equation}
    where $\mathcal{E}'_{t'}$ are spacelike surfaces orthogonal to $\partial_{t'}$, and of course $g'$ is the determinant of the metric in $(t',\bm{x}')$ coordinates.

    
    We can compare the covariant prescription \eqref{uFullCov} with the one made in Eq. \eqref{uNonCov} if, following the arguments of \cite{eduardo}, we make the physically reasonable assumption that in the reference frame of the detector it is possible to separate the spacetime smearing function as $\Lambda(\bar{ \mathsf x}) = \chi(\tau)f(\bm{\bar x})$, where $\chi(\tau)$ and $f(\bm{\bar x})$ are the switching and smearing functions from the previous sections~\footnote{This choice is made because one usually has good reasons to privilege the centre of mass of the detector, at the expense of its neighbouring points. For instance, in a realistic scenario in which the detector is an atom and the interaction with the field excites transitions between atomic energy levels, the well-known functional form of the atomic level wavefunctions---which give rise to the smearing functions---are given by solutions to the Schrodinger equation in the centre of mass of the atom \cite{eduardo}}. In this case, the the fully covariant time evolution operator from \eqref{uFullCov} taken in the reference frame of the detector can be rewritten as
    \begin{equation}\label{uCov}
        \hat{\mathcal{U}} = \mathcal{T} \mathrm{exp}\left(-\ii\lambda\int_{\mathbb{R}} \!\!\dd\tau\int_{\Sigma_{\tau}}\!\!\dd^n \bar{\bm x} \sqrt{-\bar{g}}\chi(\tau)f(\bar{\bm x})\hat{\mu}(\tau)\hat{\phi}(\bar{\mathsf{x}}) \right).
    \end{equation}
   If we compare this expression with the non-covariant prescription of Eq. \eqref{uNonCov}, we see that time evolution operators in the two prescriptions are different, i.e.,
    \begin{equation}
        \hat{U}\neq \hat{\mathcal{U}}.
    \end{equation}
    The fundamental difference between \eqref{uNonCov} and \eqref{uCov} is that the first arises from assigning a simultaneous interaction between detector and field at each \mbox{spacelike} slice associated to the centre of mass' trajectory. This is the main reason why $\hat{U}$ explicitly breaks covariance: in the presence of curvature or acceleration the four-velocity of the centre of mass of the detector $\partial_\tau$ is not the four-velocity of every point of the smeared detector moving rigidly. The operator $\hat{\mathcal{U}}$, on the other hand, takes into account that the detector has to move rigidly and therefore every point in the smearing moves with four-velocities tangent to their corresponding trajectories, thus being consistent with the view of it as a congruence of coherent subsystems, each of them mediating a point-wise interaction with the field across the smearing. 
    
    
    We can quantify the difference between the covariant model and the privileged-detector-frame one. There is, in essence, only one difference between the integrals: the fact that the first one depends on the determinant of the full metric in Fermi normal coordinates, while the other one depends on the metric in each one of the sheaves $\Sigma_\tau$. In appendix \ref{expansion} we use the results from \cite{poisson}  to write an expansion for $\sqrt{-\bar{g}}$ in terms of $\sqrt{\bar{g}_\Sigma}$ in the spatial extension of the detector. We obtained
    \begin{equation}
        \sqrt{-\bar{g}} = \sqrt{\bar{g}_\Sigma}\left(1+a_i\bar{x}^i + \frac{1}{2}R_{\tau i \tau j}\bar{x}^i\bar{x}^j\right) +\mathcal{O}(r^3),
    \end{equation}
    where $r = \sqrt{\bar{x}_i\bar{x}^i}$, $\mathsf{a} = a^i\mathsf{e}_i$ is the proper acceleration of the detector and $R_{\tau i\tau j}$ are the corresponding components of the curvature tensor in Fermi normal coordinates evaluated at the centre of mass of the detector $\mathsf{z}(\tau)$. It is worth mentioning that in such coordinates, the norm of the vector $\bm{\bar x} = (\bar{x}^i)$ corresponds to the proper distance in the sheaves $\Sigma_\tau$, so that the parameter of expansion corresponds to the physical distance from the detector to the point one integrates. That suggests, as we will discuss below, that for small enough detectors the two approaches are equally valid.
    
To study this difference in detail, let us evaluate the argument of the exponent of \eqref{uFullCov} expanded as powers of the local curvature and the proper acceleration at the centre of mass of the detector:
    \begin{align}
        &\int_{\mathbb{R}} \!\dd t \:\hat{\mathcal{H}}^t_I =\int_{\mathbb{R}} \!\dd \tau \:\hat{\mathcal{H}}^\tau_I = \int_{\mathbb{R}}\!\dd\tau \int_{\Sigma_\tau}\!\! \!\dd^n\bar{\bm x} \sqrt{-\bar{g}}\: \hat h_I(\bar{\mathsf{x}})\label{theEnd}\\
        &= \int_{\mathbb{R}} \!\dd\tau \int_{\Sigma_\tau}\!\!\! \dd^n \bar{\bm x} \sqrt{\bar{g}_\Sigma}\left(1+a_i\bar{x}^i + \frac{1}{2}R_{\tau i \tau j}\bar{x}^i\bar{x}^j\right)\hat h_I\nonumber + \mathcal{O}(r^3) \\
        &= \int_{\mathbb{R}} \!\dd\tau\left(\hat H_I^\tau(\tau) + a_i {\hat{H}^{\tau^{i}}_I}(\tau) + \frac{1}{2}R_{\tau i \tau j} {\hat{H}^{\tau^{ij}}_I}(\tau)\right) + \mathcal{O}(r^3).\nonumber
    \end{align}
    This equality implies the following equality between the integrands:
    \begin{equation}\label{hTauExpansion}
        \hat{\mathcal{H}}^\tau_I = \hat H_I^\tau(\tau) + a_i {\hat{H}^{\tau^{i}}_I}(\tau) + \frac{1}{2}R_{\tau i \tau j} {\hat{H}^{\tau^{ij}}_I}(\tau) + \mathcal{O}(r^3).
    \end{equation}
    Here $H_I(\tau),H_I^i(\tau)$ and $H_I^{ij}(\tau)$ are the monopole, dipole and quadrupole moments of the Hamiltonian weight in the sheaves, defined as
    \begin{align}
        \hat H^\tau_I(\tau) &\coloneqq \int_{\Sigma_\tau}\!\! \!\dd^n \bar{\bm x} \sqrt{\bar{g}_\Sigma}\: \hat h_I(\bar{x}),\\
         {\hat H_I^{\tau^{i}}}(\tau) &\coloneqq \int_{\Sigma_\tau}\!\! \!\dd^n \bar{\bm x} \sqrt{\bar{g}_\Sigma}\: \bar{x}^i \hat h_I(\bar{x}),\\
         {\hat H_I^{\tau^{ij}}}(\tau) &\coloneqq \int_{\Sigma_\tau}\!\! \! \dd^n \bar{\bm x} \sqrt{\bar{g}_\Sigma}\: \bar{x}^i\bar{x}^j \hat h_I(\bar{x}).
    \end{align}
    Notice that these integrals depend only on properties of the Hamiltonian in each one of the sheaves. Moreover, the first term is exactly the Hamiltonian of the non-covariant description in the detector's reference frame (Eq. \eqref{Prescription1}) if we work with the assumption that \mbox{$\Lambda(\bar{\mathsf x}) = \chi(\tau)f(\bm{\bar x})$}, that is, the physically reasonable assumption that the spacetime smearing factors into a switching function and a smearing function in the centre of mass frame of the detector \cite{eduardo}.
    
    Equation \eqref{theEnd} then allows us to compare the two terms in the exponents of $\hat{U}$ and $\hat{\mathcal{U}}$. From \eqref{hTauHT},  $\hat H^t_I$ is given by
    \begin{equation}
        \hat H^t_I = \dv{\tau}{t} \hat H^\tau_I.
        \label{thisone}
    \end{equation}
    Using equations \eqref{hTauExpansion} and \eqref{thisone}, we arrive to a  relationship between the Hamiltonians generating translations with respect to lab frame time in the non-covariant model (i.e, $\hat H^t_I$) and in the covariant model (i.e., $\hat{\mathcal{H}}^t_I$ defined according to \eqref{hCovGeneral}):  
    \begin{equation}\label{theComparison}
        \hat H^t_I =  \hat{\mathcal{H}}^t_I - \dv{\tau}{t} \qty( a_i {\hat{H}_I^{\tau^{i}}} + \frac{1}{2}R_{\tau i \tau j}{\hat{H}_I^{\tau^{ij}}})+\mathcal{O}(r^3).
    \end{equation}
    Notice that the first term  is what would covariantly be defined as the Hamiltonian in the quantization reference frame, while we get a correction that transforms as in equation \eqref{hTauHT}. We see that the first correction term (the dipole term) appears for any non-inertial motion of the detector, even in flat spacetime. The second term (the quadrupole term) is the first correction due to spacetime curvature.

   One can wonder whether the error incurred when considering the non-covariant model is relevant in the regimes where these models are usually employed. 
   
   To get an idea of the orders of magnitude introduced by the covariance correcting terms in \eqref{theComparison} for usual experimental settings, let us  assume that the detector models an atom, so that its typical size is of the order of the Bohr radius $2a_0\approx 10^{-10}$m. In this case, the order of magnitude for the acceleration of the detector so that the dipole term becomes relevant is when we have accelerations $a \approx 10^{26}g$, where $g$ the surface gravity of the Earth. This is far from the values achieved in experimental settings with atoms and actually beyond  the acceleration levels where the electromagnetic interaction is not enough to keep the rigidity of the atom. Compare this with the acceleration  needed to maintain a circular orbit in the Large Hadron Collider, which is around $10^{13}g$. For the quadrupole order term to contribute, we need a gravitational field such that the radius of curvature is of the order of the size of the detector, which would not even happen for any macroscopic black hole scenarios (outside or on the event horizon). Notice that the radius of curvature on the horizon of a solar mass black hole is of the order of kilometers, that is, again $13$ orders of magnitude away from the typical size for such detectors. This means that for practical purposes and experimental settings, smeared detectors can be modelled in their own reference frames in the way that has been traditionally done in the literature, even though this prescription breaks covariance in the ways discussed above. However, there are some scenarios where switching to a different frame is required (e.g. multiple detectors in different states of motion) and in that case having a covariant formulation is fundamental.
    
    \section{Conclusions}

    In this paper we have discussed  two different ways of handling smeared Unruh-DeWitt detectors in curved spacetimes.
    
    The first description that we have studied is usually the one found in literature, (e.g. \cite{Schlicht,Jorma,eduardo}), which is adapted to  the perspective of the reference frame of the detector. We have discussed how this prescription is not fully satisfactory for smeared detectors in curved spacetimes since it explicitly breaks covariance. We have shown how the use of this model for curved spacetimes or accelerated trajectories would not yield the same predictions (for example, detector transition probabilities) in all reference frames.
    
    Then we presented a fully covariant prescription for a detector-field interaction that naturally captures the expected covariant behaviour of the theory. By doing so we have been able to give the form of the interaction in the Fermi-Walker frame of the detector's centre of mass, where first principle arguments dictate the form of the Hamiltonian weight if we want the model to capture the features of the light-matter interaction.
    
    While being able to reproduce the same general physical results of the non-covariant model when the non-covariant approach is a good approximation (e.g., the two approaches are equivalent for pointlike detectors), the fully covariant approach removes any frame dependence of the physical predictions from the model, something essential if we discuss relativistic setups.
    
    We have also analyzed what is the error incurred when using the non-covariant approach instead of the fully covariant one. We have explicitly computed the difference between  the Hamiltonians in the fully covariant case and  in the non-covariant case. We have done so in two cases: a) when they generate time translations  with respect to the detector's centre of mass proper time (that is, the difference $\hat{\mathcal{H}}^\tau_I-\hat{H}^\tau_I$)  and b) when they generate translations with respect to  the lab's time (that is, the difference $\hat{\mathcal{H}}^t_I-\hat{H}^t_I$).
    
    We have proven that the difference between the two approaches can be expressed in terms of a power series on the smearing lenghtscale of the detector, and estimated the magnitude of the acceleration for the detector or the curvature of spacetime that would be needed for the non-covariant model to yield significantly different results than the covariant one. We found that until we reach accelerations, or gravitational fields of  order of  $10^{26}g$ for an atomic sized detector the two approaches give approximately the same results in the detector's proper frame. These accelerations are way below those where the Unruh effect or other similar relativistic  phenomena  should be visible (see e.g., \cite{ChenTaj,BerryPh,Matsas,ScullyPage}) and therefore we conclude that for non-extreme spacetime curvature or accelerations the non-covariant model can be a good approximation to the covariant one.
    
    However in scenarios where several smeared detectors have to be considered, and the causality of the predictions is relevant for any conclusions extracted  (e.g., in entanglement harvesting setups or in quantum communication), we have to restrict to the fully covariant formulation, as different detectors in a curved spacetime (or in different states of motion in flat spacetime) would have their Hamiltonians prescribed in different reference frames.
    
\section{Acknowledgements}

The authors thank Jonas Neuser for insightful discussions. E.M-M acknowledges the support of the NSERC Discovery program as well as his Ontario Early Researcher Award. 
   
    
    

\appendix
\onecolumngrid
    \section{Volume element in Fermi normal coordinates}\label{expansion}
    From \cite{poisson} the metric $\bar{g}$ in Fermi normal coordinates around the curve $\mathsf z(\tau)$ can be expressed as
    \begin{align}
        \bar{g}_{\tau\tau} &= -(1+2a_i \bar{x}^i+(a_i \bar{x}^i)^2+ R_{\tau i \tau j}\bar{x}^i\bar{x}^j) + \mathcal{O}(r^3),\quad
        \bar{g}_{\tau i} = -\frac{2}{3} R_{\tau kij}\bar{x}^k\bar{x}^j + \mathcal{O}(r^3),\quad
        \bar{g}_{ij} = \delta_{ij}-\frac{1}{3}R_{ikjl}\bar{x}^k\bar{x}^l + \mathcal{O}(r^3),
    \end{align}
    where $\bar{x}^i$ are the coordinates in each of the sheaves, ${r^2 = \sum_i (\bar{x}^i)^2}$ is the geodesic distance between the curve and each of the points in $\Sigma_\tau$, $R_{\alpha\beta\mu\nu}$ is the Riemann tensor evaluated at $\mathsf z(\tau)$ in Fermi normal coordinates, and $\mathsf a(\tau)$ is the proper acceleration of $\mathsf z(\tau)$.
    
    From that we can calculate the determinant of the metric $\det (\bar{g})$ in such a coordinate system in terms of the determinant of the spatial part of it, $\det (\bar{g}_{ij})$. Using the following expression for the determinant we can follow up with the calculations:
    \begin{align}
        \det \bar{g} &= \epsilon^{\alpha\beta\mu\nu}\bar{g}_{\alpha \tau}\bar{g}_{\beta 1}\bar{g}_{\mu 2}\bar{g}_{\nu 3} \nonumber\\
        &= \epsilon^{\tau ijk}\bar{g}_{\tau\tau}\bar{g}_{i1}\bar{g}_{j2}\bar{g}_{k3}-\epsilon^{\tau ijk}\bar{g}_{\tau i}\bar{g}_{\tau 1}\bar{g}_{j2}\bar{g}_{k3}
        +\epsilon^{\tau ijk}\bar{g}_{\tau j}\bar{g}_{i1}\bar{g}_{\tau 2}\bar{g}_{k3}-\epsilon^{\tau ijk}\bar{g}_{\tau k}\bar{g}_{i1}\bar{g}_{j2}\bar{g}_{\tau 3}\nonumber\\ 
        &= \bar{g}_{\tau\tau}\det(\bar{g}_{ij}) - \frac{4}{9}\epsilon^{\tau i j k} R_{\tau aib}\bar{x}^a\bar{x}^bR_{\tau c1d}\bar{x}^c\bar{x}^d\left(\delta_{j2}-\frac{1}{3}R_{je2f}\bar{x}^e\bar{x}^f\right)\left(\delta_{k3}-\frac{1}{3}R_{km3n}\bar{x}^m\bar{x}^n\right)\nonumber\\
        &\textrm{\:\:\:\:\:\:\:\:\:\:\:\:\:\:\:\:\:\:\:\:\:\:\:\:\:\:\:\:\:}  +\frac{4}{9}\epsilon^{\tau i j k} R_{\tau ajb}\bar{x}^a\bar{x}^bR_{\tau c2d}\bar{x}^c\bar{x}^d\left(\delta_{i1}-\frac{1}{3}R_{ie1f}\bar{x}^e\bar{x}^f\right)\left(\delta_{k3}-\frac{1}{3}R_{km3n}\bar{x}^m\bar{x}^n\right)\nonumber\\
        &\textrm{\:\:\:\:\:\:\:\:\:\:\:\:\:\:\:\:\:\:\:\:\:\:\:\:\:\:\:\:\:}  -\frac{4}{9}\epsilon^{\tau i j k} R_{\tau akb}\bar{x}^a\bar{x}^bR_{\tau c3d}\bar{x}^c\bar{x}^d\left(\delta_{j2}-\frac{1}{3}R_{je2f}\bar{x}^e\bar{x}^f\right)\left(\delta_{i1}-\frac{1}{3}R_{im1n}\bar{x}^m\bar{x}^n\right)+\mathcal{O}(r^5)\nonumber\\
        & =\bar{g}_{\tau\tau} \det(\bar{g}_{ij})- \frac{4}{9}\epsilon^{\tau i 2 3} R_{\tau aib}R_{\tau c1d}\bar{x}^a\bar{x}^b\bar{x}^c\bar{x}^d+\frac{4}{9}\epsilon^{\tau 1 j 3} R_{\tau ajb}R_{\tau c2d}\bar{x}^a\bar{x}^b\bar{x}^c\bar{x}^d -\frac{4}{9}\epsilon^{\tau 1 2 k} R_{\tau akb}R_{\tau c3d}\bar{x}^a\bar{x}^b\bar{x}^c\bar{x}^d+\mathcal{O}(r^5)\nonumber\\
        &= \bar{g}_{\tau\tau}\det(\bar{g}_{ij}) -\frac{4}{9}\left(\epsilon^{\tau i 2 3} R_{\tau aib}R_{\tau c1d}-\epsilon^{\tau 1 j 3} R_{\tau ajb}R_{\tau c2d} +\epsilon^{\tau 1 2 k} R_{\tau akb}R_{\tau c3d}\right)\bar{x}^a\bar{x}^b\bar{x}^c\bar{x}^d+\mathcal{O}(r^5)\nonumber\\
        &= \bar{g}_{\tau\tau}\det(\bar{g}_{ij}) -\frac{4}{9}\left(\epsilon^{\tau 1 2 3} R_{\tau a1b}R_{\tau c1d}-\epsilon^{\tau 1 2 3} R_{\tau a2b}R_{\tau c2d} +\epsilon^{\tau 1 2 3} R_{\tau a3b}R_{\tau c3d}\right)\bar{x}^a\bar{x}^b\bar{x}^c\bar{x}^d+\mathcal{O}(r^5)\nonumber\\
        &= \bar{g}_{\tau\tau}\det(\bar{g}_{ij}) -\frac{4}{9}\left( R_{\tau a1b}R_{\tau c1d}- R_{\tau a2b}R_{\tau c2d} +R_{\tau a3b}R_{\tau c3d}\right)\bar{x}^a\bar{x}^b\bar{x}^c\bar{x}^d+\mathcal{O}(r^5)\nonumber\\
        &= -\det(\bar{g}_{ij})-\left(2a_i\bar{x}^i+(a_i\bar{x}^i)^2+R_{\tau i\tau j}\bar{x}^i\bar{x}^j\right)\det(\bar{g}_{ij})+\mathcal{O}(r^4).
    \end{align}
   This yields:
    \begin{align}\label{sqrtGIJ}
        \begin{split}
        \sqrt{-\bar{g}} &= \sqrt{\bar{g}_\Sigma}\sqrt{1+2a_i\bar{x}^i+(a_i\bar{x}^i)^2+R_{\tau i \tau j}\bar{x}^i\bar{x}^j+\mathcal{O}(r^4)} \\
        &=\sqrt{\bar{g}_\Sigma}\left(1+a_i\bar{x}^i +\frac{1}{2}(a_i\bar{x}^i)^2 +\frac{1}{2}R_{\tau i \tau j}\bar{x}^i\bar{x}^j-\frac{1}{8}(2a_i\bar{x}^i)^2+\mathcal{O}(r^3)\right)\\
        &= \sqrt{\bar{g}_\Sigma}\left(1+a_i\bar{x}^i + \frac{1}{2}R_{\tau i \tau j}\bar{x}^i\bar{x}^j\right) + \mathcal{O}(r^3)
        \end{split}
    \end{align}
    We also can calculate the determinant of $\bar{g}_{ij}$ in terms of the curvature tensor:
    \begin{align}
        \begin{split}
        \bar{g}_\Sigma &= \epsilon^{ijk}\bar{g}_{i1}\bar{g}_{j2}\bar{g}_{k3} = \epsilon^{ijk}\left(\delta_{i1}-\frac{1}{3}R_{im1n}\bar{x}^m\bar{x}^n\right)\left(\delta_{j2}-\frac{1}{3}R_{jm2n}\bar{x}^m\bar{x}^n\right)\left(\delta_{k3}-\frac{1}{3}R_{km3n}\bar{x}^m\bar{x}^n\right)\\
        &= 1-\frac{1}{3}(\epsilon^{i23}R_{im1n}+\epsilon^{1j3}R_{jm2n}+\epsilon^{12k}R_{km3n})\bar{x}^m\bar{x}^n+\mathcal{O}(r^4)\\
        &= 1-\frac{1}{3}(R_{1i1j}+R_{2i2j}+R_{3i3j})\bar{x}^i\bar{x}^j+\mathcal{O}(r^4).
        \end{split}
    \end{align}
    Giving us: $\displaystyle{
        \sqrt{\bar{g}_\Sigma} = 1-\frac{1}{6}(R_{1i1j}+R_{2i2j}+R_{3i3j})\bar{x}^i\bar{x}^j + \mathcal{O}(r^4).}$
    In the end, replacing this in the expression \eqref{sqrtGIJ}, we get:
    \begin{align}
        \sqrt{-g} = 1 + a_i\bar{x}^i +\frac{1}{2}\left(R_{\tau i \tau j}-\frac{1}{3}(R_{1 i 1 j}+R_{2i2j}+R_{3i3j})\right) \bar{x}^i\bar{x}^j+\mathcal{O}(r^3)
    \end{align}

\twocolumngrid

\bibliography{references}

\end{document}